%% file: 0_Main.tex
\documentclass{article}
\PassOptionsToPackage{numbers}{natbib}

\usepackage[final]{styles/neurips_2020}

\usepackage[utf8]{inputenc} 
\usepackage[T1]{fontenc}    
\usepackage{hyperref}
\hypersetup{hidelinks}
\usepackage{url}            
\usepackage{booktabs}       
\usepackage{nicefrac}       
\usepackage{microtype}      
\usepackage{graphicx}
\usepackage{amsfonts}       
\usepackage{amsmath,amssymb} 
\usepackage{mathtools}
\usepackage{bm}
\usepackage{todonotes}
\usepackage{soul}
\usepackage{sidecap}
\usepackage{xspace}
\makeatletter
 \if@todonotes@disabled
 \newcommand{\hlnote}[2]{#1}
 \else
 \newcommand{\hlnote}[2]{\todo{#2}\texthl{#1}}
 \fi
 \makeatother
\usepackage{url}
\usepackage[inline]{enumitem}

\usepackage{floatrow}
\usepackage{multirow}
\usepackage{comment}
\usepackage{subcaption}
\newfloatcommand{capbtabbox}{table}[][\FBwidth]

\usepackage{authblk}

\input{ourcommands}
\input{macros}

\title{Listening to Sounds of Silence for Speech Denoising}

\author[1]{Ruilin Xu}
\author[1]{Rundi Wu}
\author[2]{Yuko Ishiwaka}
\author[1]{Carl Vondrick}
\author[1]{Changxi Zheng}

\affil[1]{Columbia University, New York, USA}
\affil[2]{SoftBank Group Corp., Tokyo, Japan}

\begin{document}

\maketitle

\vspace{-3mm}
\begin{abstract}
\vspace{-1.2mm}
We introduce a deep learning model for speech denoising, a long-standing challenge in audio analysis
arising in numerous applications.
Our approach is based on a key observation about human speech: there is often a short
pause between each sentence or word.
In a recorded speech signal, those pauses introduce a series of time periods during which only noise 
is present.
We leverage these incidental \emph{silent intervals} to learn a model for automatic speech denoising given only mono-channel audio.
Detected silent intervals over time expose not just pure noise but its
time-varying features, allowing the model to learn noise dynamics and suppress it from the
speech signal.
Experiments on multiple datasets confirm the pivotal role of silent interval detection for
speech denoising, and our method outperforms several state-of-the-art
denoising methods, including those that accept only audio input (like ours) and those that
denoise based on audiovisual input (and hence require more information).
We also show that our method enjoys excellent generalization properties, such
as denoising spoken languages not seen during training. 
\end{abstract}

\input{1_Introduction}

\input{2_Related_Work}

\input{3_Method}

\input{4_Experiments}
\input{5_Conclusion}

\paraspace
\paragraph{Acknowledgments.}
This work was supported in part by the National Science Foundation (1717178, 1816041, 1910839, 1925157)
and SoftBank Group.

\newpage

\input{6_Broader}

\bibliographystyle{abbrvnat}
\bibliography{References}

\newpage
\setcounter{page}{1}
\appendix
\input{app}

\end{document}

%% file: ourcommands.tex
\usepackage{color}
\definecolor{green}{rgb}{0.23, 0.7, 0.45}
\definecolor{orange}{rgb}{0.8, 0.6, 0.2}
\definecolor{red}{rgb}{0.88, 0.33, 0.33}
\definecolor{teal}{rgb}{0.2, 0.54, 0.65}
\definecolor{blue}{rgb}{0.3, 0.62, 0.88}
\definecolor{purple}{rgb}{0.47, 0.41, 0.68}
\definecolor{pink}{rgb}{0.88, 0.55, 0.67}
\definecolor{saffron}{rgb}{0.88, 0.74, 0.16}
\definecolor{turquoise}{rgb}{0.0, 0.5, 0.5}
\definecolor{black}{rgb}{0.0, 0.0, 0.0}

\newcommand{\rw}[1]{{\color{red}\{\textbf{Rundi:} #1\}}}
\newcommand{\hx}[1]{{\color{green}\{\textbf{Henry:} #1\}}}
\newcommand{\cv}[1]{{\color{blue}\{\textbf{Carl:} #1\}}}

\newcommand{\sidModule}{\mathsf{D}}
\newcommand{\nRecoverModule}{\mathsf{N}}
\newcommand{\nRemoveModule}{\mathsf{R}}

\newcommand{\noisyAudio}{\bm{x}}
\newcommand{\cleanAudio}{\bm{x^*}}

\newcommand{\noiseProfileAudio}{\tilde{\bm{x}}}
\newcommand{\gtnoiseProfileAudio}{\mathbf{\widetilde x_p^{GT}}}

\newcommand{\noisyAudioSTFT}{\bm{s_{x}}}
\newcommand{\cleanAudioSTFT}{\bm{s}^*_{\bm{x}}}
\newcommand{\outputSTFT}{\bm{s_{x}^*}}
\newcommand{\noiseAudioSTFT}{\bm{s}^*_{\bm{n}}}
\newcommand{\noiseProfileAudioSTFT}{\bm{s}_{\tilde{\bm{x}}}}
\newcommand{\gtnoiseProfileAudioSTFT}{\mathbf{S_{\gtnoiseProfileAudio}}}

\newcommand{\estimatedMask}{\bm{m}(\bm{x})}
\newcommand{\silentIntervalMask}{\bm{m}^*_{\bm{x}}}

\newcommand{\complexMask}{\mathbf{c}}

\newcommand{\lossBCE}{\mathcal{L}_1}
\newcommand{\lossMSE}{\mathcal{L}_0}

\newcommand{\DEM}{\texttt{DEMAND}\xspace}
\newcommand{\AS}{\texttt{AudioSet}\xspace}
\newcommand{\GTSI}{\texttt{Ours-GTSI}\xspace}
\newcommand{\AVSP}{\texttt{AVSPEECH}\xspace}

%% file: macros.tex
\newcommand{\fixme}[1]{\textbf{\color{red}[#1]}}
\newcommand{\rev}[1]{{\color{red}#1}} 

\newcommand{\secprespace}{\vspace{-2.2mm}}
\newcommand{\secspace}{\vspace{-2.2mm}}
\newcommand{\paraspace}{\vspace{-1.2mm}}

\newcommand{\figref}[1]{Fig.~\ref{fig:#1}}
\newcommand{\tabref}[1]{Table~\ref{tab:#1}}
\newcommand{\secref}[1]{Sec.~\ref{sec:#1}}

\newcommand{\appref}[1]{Appendix~\ref{sec:#1}}

\newcommand{\eq}[1]{\eqref{eq:#1}}
\newcommand{\norm}[1]{\left\lVert#1\right\rVert}

\makeatletter
\renewcommand{\paragraph}{%
  \@startsection{paragraph}{4}%
  {\z@}{0.60ex \@plus 1ex \@minus .15ex}{-1em}%
  {\normalfont\normalsize\bfseries}%
}
\makeatother

%% file: 1_Introduction.tex
\vspace{-0.5mm}
\secprespace
\section{Introduction}\label{sec:intro}
\secspace




Noise is everywhere.  When we listen to someone speak, the audio signals we
receive are never pure and clean, always contaminated by all kinds of
noises---cars passing by, spinning fans in an air conditioner, barking dogs,
music from a loudspeaker, and so forth. 
To a large extent, people in a conversation can effortlessly filter out these noises~\cite{Kell2019}. 
In the same vein, numerous applications, ranging from cellular communications to human-robot interaction,
rely on speech denoising algorithms as a fundamental building block.


Despite its vital importance, algorithmic speech denoising remains a grand challenge.
Provided an input audio signal, speech denoising aims to separate the foreground (speech) signal
from its additive background noise. This separation problem is inherently ill-posed.
Classic approaches such as spectral subtraction~\cite{boll, Weiss, 1170788, Yang2005SpectralSS, sg} and Wiener filtering~\cite{543199, 1163086}
conduct audio denoising in the spectral domain, and they are typically restricted to
stationary or quasi-stationary noise.
In recent years, the advance of deep neural networks has also inspired their use in audio denoising.
While outperforming the classic denoising approaches, existing neural-network-based approaches
use network structures developed for general audio processing tasks~\cite{Maas2012RecurrentNN, Valentini-Botinhao+2016, 101007} or borrowed 
from other areas such as computer vision~\cite{10112111358887, Gabbay2017, Afouras2018, Hou2018, m2019av} and generative adversarial networks~\cite{Pascual2017, Pascual2019}. 
Nevertheless, beyond reusing well-developed network models as a black box, a fundamental question remains:
\emph{
What natural structures of speech can we leverage to mold network architectures for better performance on speech denoising?
}




\secprespace
\subsection{Key insight: time distribution of silent intervals}\label{sec:SI}
\secspace
Motivated by this question,
we revisit one of the most widely used audio denoising methods in 
practice, namely the spectral subtraction method~\cite{boll, Weiss, 1170788, Yang2005SpectralSS, sg}.
Implemented in many commercial software such as Adobe Audition~\cite{audition},
this classical method requires the user to specify a time interval during which the foreground signal
is absent. We call such an interval a \emph{silent interval}.
A silent interval is a time window that exposes pure noise.
The algorithm then learns from the silent interval the noise characteristics, which are in turn
used to suppress the additive noise of the entire input signal (through subtraction in the spectral domain).

Yet, the spectral subtraction method suffers from two major shortcomings: i) it
requires user specification of a silent interval, that is, not fully automatic; and ii)
the single silent interval, although undemanding for the user, is insufficient
{in presence} of \emph{nonstationary} noise---for example, a background music.
Ubiquitous in daily life, nonstationary noise has time-varying spectral
features. 
The single silent interval reveals the noise spectral features
only in that particular time span, thus inadequate for denoising the entire
input signal. The success of spectral subtraction pivots on the concept of silent interval;
so do its shortcomings.



\begin{figure*}[t]
\centering
\includegraphics[width=0.94\linewidth]{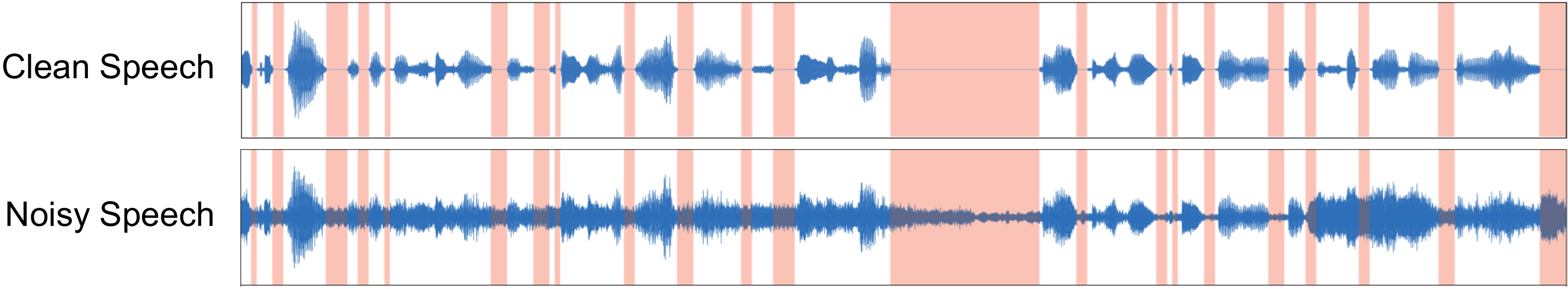}
\vspace{-2.5mm}
\caption{{\bf Silent intervals over time.} 
\textbf{(top)} A speech signal has many natural pauses.
Without any noise, these pauses are exhibited as silent intervals (highlighted in red).
\textbf{(bottom)} However, most speech signals are contaminated by noise.
Even with mild noise, silent intervals become overwhelmed and hard to detect.
If robustly detected, silent intervals can help to reveal the noise profile over time.
\label{fig:si}}
\end{figure*}

In this paper, we introduce a deep network for speech denoising that tightly integrates silent intervals, and thereby overcomes many of the limitations of classical approaches.
Our goal is not just to identify 
a single silent interval, but to find as many as possible silent intervals over time.
Indeed, silent intervals in speech appear in abundance:
psycholinguistic studies have shown that
there is almost always a pause after each sentence and even each word in speech~\cite{rochester1973significance,fors2015production}.
Each pause, however short, provides a silent interval revealing noise characteristics local in time.
All together, these silent intervals assemble a time-varying picture of background noise,
allowing the neural network to better denoise speech signals,
even {in presence} of nonstationary noise (see \figref{si}).

In short, to interleave neural networks with established denoising pipelines, we propose a network structure consisting of three major components (see \figref{network}): 
\textbf{i)} one  
dedicated to silent interval detection, \textbf{ii)} another that aims to estimate the full noise from
those revealed in silent intervals, akin to an inpainting process in computer vision~\cite{Iizuka2017completion},
and \textbf{iii)} yet another for cleaning up the input signal.

\textbf{Summary of results.}
Our neural-network-based denoising model accepts a single channel of audio signal
and outputs the cleaned-up signal.
Unlike some of the recent denoising methods that take as input audiovisual signals (i.e.,
both audio and video footage), our method can be applied in a wider range of scenarios (e.g., in cellular communication).
We conducted extensive experiments, including ablation studies to show the efficacy 
of our network components and comparisons to several state-of-the-art denoising methods.
{We also evaluate our method under various signal-to-noise ratios---even under strong noise levels
that are not tested against in previous methods.}
We show that, under a variety of denoising metrics, our method consistently outperforms those methods, including
those that accept only audio input (like ours) and those that denoise based on audiovisual input.

The pivotal role of silent intervals for speech denoising is further confirmed by a few key results.
Even without supervising on silent interval detection, the ability to detect silent intervals 
naturally emerges in our network.
Moreover, while our model is trained on English speech only, with no additional training it can be readily used to
denoise speech in other languages (such as Chinese, Japanese, and Korean). 
Please refer to the supplementary materials for listening to our denoising results.

%% file: 2_Related_Work.tex
\secprespace
\section{Related Work}
\secspace

\begin{figure}[t]
\begin{center}
\includegraphics[width=0.98\linewidth,keepaspectratio]{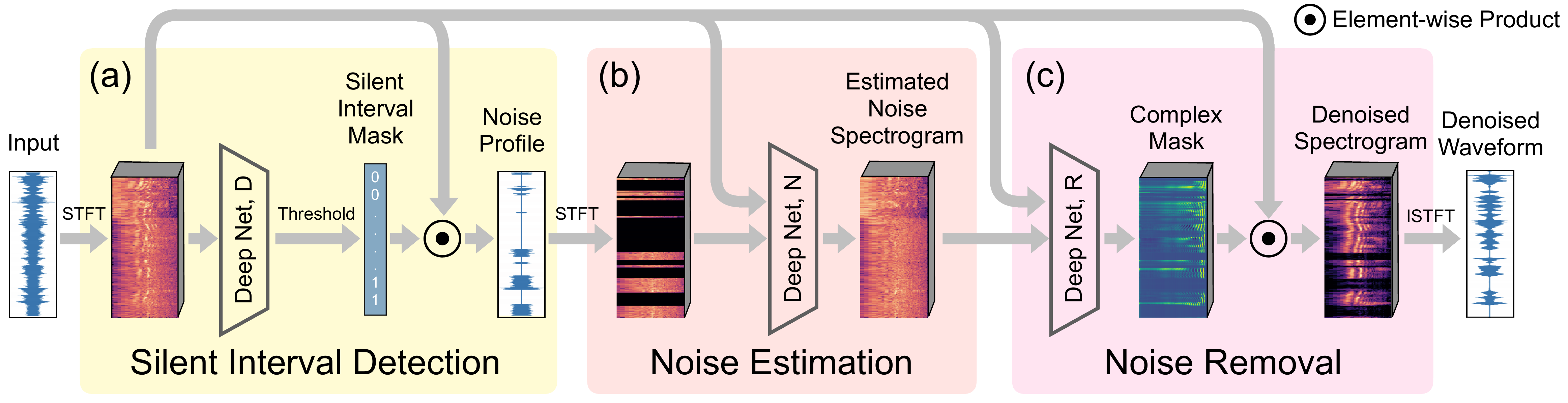}
\end{center}
\vspace{-4.8mm}
\caption{{\bf Our audio denoise network.} 
Our model has three components: \textbf{(a)} one that detects silent intervals over time,
and outputs a noise profile observed from detected silent intervals;
\textbf{(b)} another that estimates the full noise profile, and \textbf{(c)} yet another
that cleans up the input signal.
\label{fig:network}}
\end{figure}

\paragraph{Speech denoising.} Speech denoising~\cite{se} is a fundamental problem studied over several decades. 
Spectral subtraction ~\cite{boll, Weiss, 1170788, Yang2005SpectralSS, sg} estimates the clean signal spectrum by subtracting an estimate of the noise spectrum from the noisy speech spectrum. 
This classic method was followed by spectrogram factorization methods~\cite{6784107}.
Wiener filtering~\cite{543199, 1163086} derives the enhanced signal by optimizing the mean-square error. 
Other methods exploit pauses in speech, forming segments of low acoustic energy where noise statistics can be more accurately measured \cite{Doblinger95computationallyefficient, 928915,10.1155/ASP.2005.2954, 1164550, 1325983, 1223596, 988717}.
Statistical model-based methods~\cite{168664, Hirsch1995NoiseET} and subspace algorithms~\cite{101016, 397090} are also studied.

Applying neural networks to audio denoising dates back to the 80s~\cite{196643, 1326090}. With increased computing power, deep neural networks are often used~\cite{6665000, Xu2015MultiobjectiveLA, 6932438, Kumar2016SpeechEI}. Long short-term memory networks (LSTMs)~\cite{lstm} are able to preserve temporal context information of the audio signal~\cite{lipton2015critical}, leading to strong results \cite{Maas2012RecurrentNN, Valentini-Botinhao+2016, 101007}. Leveraging generative adversarial networks (GANs)~\cite{10.5555/2969033.2969125}, methods such as~\cite{Pascual2017, Pascual2019} have adopted GANs into the audio field and have also achieved strong performance.

Audio signal processing methods operate on either the raw waveform or the
spectrogram by Short-time Fourier Transform (STFT). Some work directly on 
waveform~\cite{Fu_2017, Pandey2018, michelashvili2019audio, 101109TASLP20192915167}, and
others use Wavenet~\cite{Oord2016aiw} for speech
denoising~\cite{Qian2017, 8462417, Germain2019}. Many other methods such as~\cite{Lu2013SpeechEB, 10555529991342999160,
6639038, 7032183, Kumar_2016, 7422753, 2017ASAJ1414705C} work on audio signal's
spectrogram, which contains both magnitude and phase information.
There are works discussing how to use the spectrogram to its best
potential~\cite{1163920, 10101622}, while one of the
disadvantages is that the inverse STFT needs to be applied. 
Meanwhile, there also exist works~\cite{6334422,
7471747, 7038277, 7178800, 7178061, 7906509, 8170014} investigating how to
overcome artifacts from time aliasing.

Speech denoising has also been studied in conjunction with computer vision due to the relations between speech and facial features~\cite{4317558}. Methods such as~\cite{10112111358887, Gabbay2017, Afouras2018, Hou2018, m2019av} utilize different network structures to enhance the audio signal to the best of their ability. Adeel et al.~\cite{Adeel2019} even utilize lip-reading to filter out the background noise of a speech.

\paraspace
\paragraph{Deep learning for other audio processing tasks.} Deep learning is
widely used for lip reading, speech recognition, speech separation, and many
audio processing or audio-related tasks, with the help of computer vision
\cite{Owens_2018,owens2016ambient,aytar2016soundnet,arandjelovic2018objects}.
Methods such as~\cite{7949073, 8265271, Owens_2016} are able to reconstruct
speech from pure facial features. Methods such as~\cite{8585066,
101007s1048901406297} take advantage of facial features to improve speech
recognition accuracy. Speech separation is one of the areas where computer vision is best leveraged. Methods such as~\cite{gabbay2017seeing,
Owens_2018, Ephrat_2018,zhao2018sound} have achieved impressive results, making
the previously impossible speech separation from a single audio signal
possible. Recently, Zhang et al.~\cite{Zhang2020Deep} proposed a new operation
called Harmonic Convolution to help networks distill audio priors, which is
shown to even further improve the quality of speech separation.

%% file: 3_Method.tex
\secprespace
\section{Learning Speech Denoising}\label{sec:method}
\secspace


We present a neural network that harnesses the time distribution
of silent intervals for speech denoising.
The input to our model is a spectrogram
of noisy speech~\cite{wyse2017audio, sasp, 4511413}, which can be 
viewed as a 2D image of size $T\times F$ with two channels,
where $T$ represents the time length of the signal and $F$ is the number of
frequency bins. The two channels store the real and imaginary parts of STFT,
respectively. After learning, the model will produce another spectrogram of the
same size as the noise suppressed.

We first train our proposed network structure in an end-to-end fashion, with only denoising supervision (\secref{training});
and it already outperforms the state-of-the-art methods that we compare against. 
Furthermore, we incorporate the supervision on silent interval detection (\secref{supervision}) and obtain
even better denoising results (see \secref{exp}).


\begin{figure}[ht]
    \centering
    \includegraphics[width=0.98\linewidth, page=1]{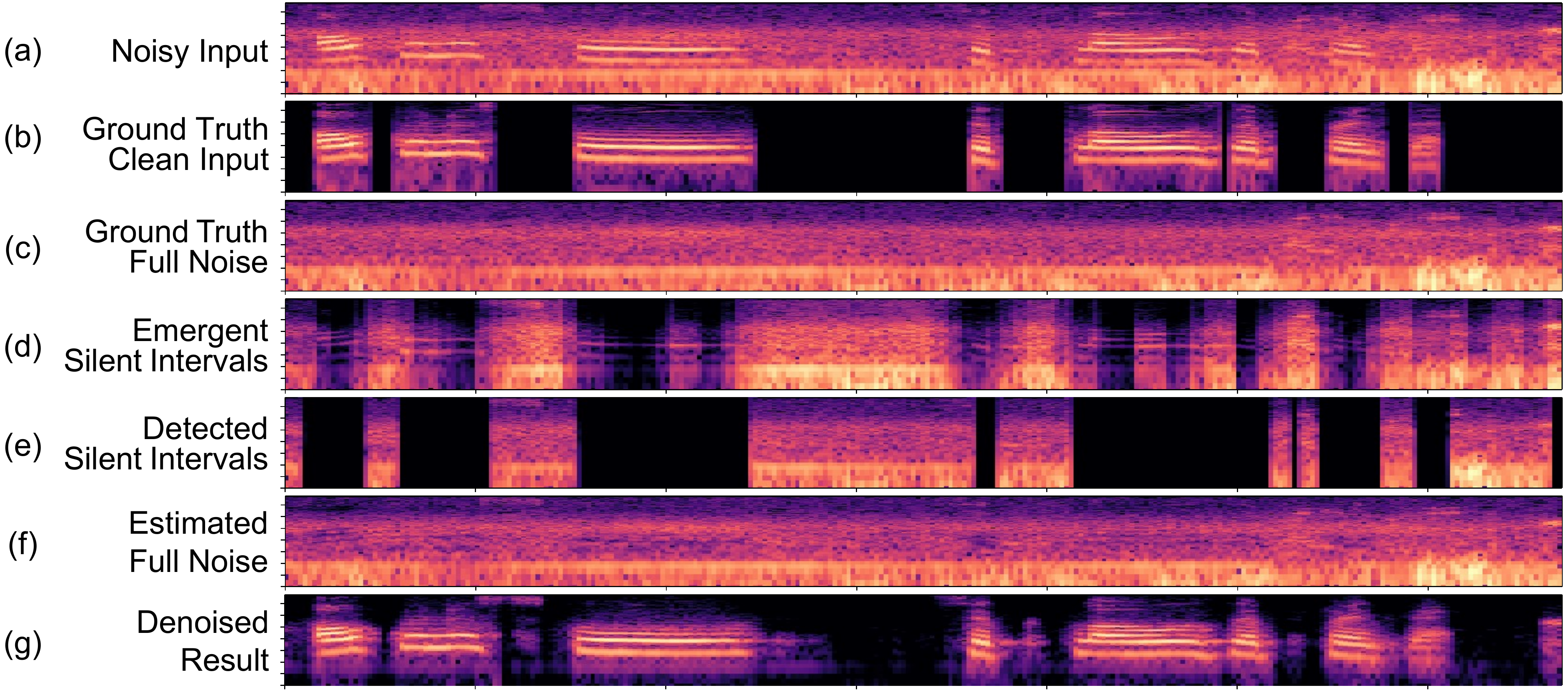}
    \vspace{-3mm}
    \caption{{\bf Example of intermediate and final results.}
    \textbf{(a)} The spectrogram of a noisy input signal,
    which is a superposition of
    a clean speech signal \textbf{(b)} and
    a noise \textbf{(c)}. 
    The \textbf{black} regions in (b) indicate ground-truth silent intervals.
    \textbf{(d)} The noise exposed by automatically emergent silent intervals,
    i.e., the output of the silent interval detection component 
    when the entire network is trained without silent interval supervision (recall \secref{training}).
    \textbf{(e)} The noise exposed by detected silent intervals,
    i.e., the output of the silent interval detection component 
    when the network is trained with silent interval supervision (recall \secref{supervision}).
    \textbf{(f)} The estimated noise profile using subfigure (a) and (e) as the input to the noise estimation component.
    \textbf{(g)} The final denoised spectrogram output.
    }
    \label{fig:results_pipeline}
\end{figure}


\secprespace
\subsection{Network structure}\label{sec:structure}
\secspace

Classic denoising algorithms work in three general stages: silent interval
specification, noise feature estimation, and noise removal. We propose to
interweave learning throughout this process: we rethink each stage
with the help of a neural network, forming a new speech denoising model. 
Since we can chain these networks together and estimate gradients, we can efficiently train the model with large-scale audio data.
Figure~\ref{fig:network} illustrates this model, which we describe below.




\paraspace
\paragraph{Silent interval detection.}
The first component is dedicated to detecting silent intervals in the input 
signal. The input to this component is the spectrogram of the input (noisy) signal $\noisyAudio$.
The spectrogram $\noisyAudioSTFT$ is first encoded by a 2D convolutional
encoder into a 2D feature map, which is in turn processed by a bidirectional
LSTM~\cite{lstm,brnn} followed by two fully-connected (FC) layers (see network details in \appref{network_details}).
The bidirectional LSTM is suitable for processing time-series features resulting from the
spectrogram~\cite{mehri2016samplernn,kalchbrenner2018efficient,Purwins_2019,Ephrat_2018}, 
and the FC layers are applied to the features of each time sample to accommodate variable length input.
The output from this network component is a vector $\sidModule(\noisyAudioSTFT)$. 
Each element of $\sidModule(\noisyAudioSTFT)$ is a scalar in [0,1] (after applying the sigmoid function),
indicating a confidence score of a small time segment being silent. We choose each time segment to have
$\nicefrac{1}{30}$ second, small enough to capture short speech pauses and
large enough to {allow robust prediction}. 

The output vector $\sidModule(\noisyAudioSTFT)$ is then expanded to a longer mask, 
which we denote as $\estimatedMask$. Each element of this mask 
indicates the confidence of classifying each sample of the input signal $\noisyAudio$
as pure noise (see \figref{results_pipeline}-e).
With this mask, the noise profile $\noiseProfileAudio$ exposed by silent intervals are estimated by
an element-wise product, namely $\noiseProfileAudio\coloneqq\noisyAudio\odot\estimatedMask$.

\paraspace
\paragraph{Noise estimation.}
The signal $\noiseProfileAudio$ resulted from silent interval detection is
noise profile exposed only through a series of time windows (see
\figref{results_pipeline}-e)---but not a complete picture of the noise.  However,
since the input signal is a superposition of clean speech signal and noise,
having a complete noise profile would ease the denoising process, especially {in
presence} of nonstationary noise.  Therefore, we also estimate the entire
noise profile over time, which we do with a neural network.  

Inputs to this component include both the noisy audio signal $\noisyAudio$ and the incomplete
noise profile $\noiseProfileAudio$. Both are converted by STFT into spectrograms,
denoted as $\noisyAudioSTFT$ and $\noiseProfileAudioSTFT$, respectively.
We view the spectrograms as 2D images. And because 
the neighboring time-frequency pixels in a spectrogram are often correlated,
our goal here is conceptually akin to the image inpainting task in computer vision~\cite{Iizuka2017completion}.
To this end, we encode $\noisyAudioSTFT$ and $\noiseProfileAudioSTFT$
by two separate 2D convolutional encoders into two feature maps.
The feature maps are then concatenated in a channel-wise manner and further decoded by a
convolutional decoder to estimate the full noise spectrogram, which we
denote as $\nRecoverModule(\noisyAudioSTFT, \noiseProfileAudioSTFT)$. 
A result of this step is illustrated in \figref{results_pipeline}-f.


\paraspace
\paragraph{Noise removal.}
Lastly, we clean up the noise from the input signal $\noisyAudio$. We use a neural network $\nRemoveModule$ that takes as input both the
input audio spectrogram $\noisyAudioSTFT$ and the estimated full noise spectrogram
$\nRecoverModule(\noisyAudioSTFT, \noiseProfileAudioSTFT)$.
The two input spectrograms are processed individually by their own 2D convolutional encoders.
The two encoded feature maps are then concatenated together before passing to a bidirectional
LSTM followed by three fully connected layers (see details in \appref{network_details}).
Like other audio enhancement models~\cite{Ephrat_2018, Wang_2018, 6887314},
the output of this component is a vector with two channels which form
the real and imaginary parts of a complex ratio mask
$\complexMask\coloneqq\nRemoveModule \big(\noisyAudioSTFT,\nRecoverModule(\noisyAudioSTFT, \noiseProfileAudioSTFT) \big)$
in frequency-time domain. In other words, the mask $\complexMask$ has the same (temporal and frequency) 
dimensions as $\noisyAudioSTFT$.

In the final step, we compute the denoised spectrogram $\outputSTFT$ through
element-wise multiplication of the input audio spectrogram $\noisyAudioSTFT$
and the mask $\complexMask$ (i.e.,
$\outputSTFT=\noisyAudioSTFT\odot\complexMask$).  Finally, the cleaned-up audio
signal is obtained by applying the inverse STFT to $\outputSTFT$ (see \figref{results_pipeline}-g).


\vspace{-0.525mm}
\secprespace
\subsection{Loss functions and training}\label{sec:training}
\secspace


Since a subgradient exists at every step, we are able to train our network in an end-to-end fashion with stochastic gradient descent. We optimize the following loss function:
\begin{equation}\label{eq:loss}
\lossMSE = \mathbb{E}_{\noisyAudio \sim p(\noisyAudio)} \Big[
    \norm{ \nRecoverModule(\noisyAudioSTFT, \noiseProfileAudioSTFT) - \noiseAudioSTFT }_2 + 
\beta\norm{ \noisyAudioSTFT \odot \nRemoveModule \big(\noisyAudioSTFT, \nRecoverModule(\noisyAudioSTFT, \noiseProfileAudioSTFT) \big) - \cleanAudioSTFT }_2 \Big],
\end{equation}
where the notations $\noisyAudioSTFT$, $\noiseProfileAudioSTFT$, $\nRecoverModule(\cdot,\cdot)$,
and $\nRemoveModule(\cdot,\cdot)$ are defined in \secref{structure};
$\cleanAudioSTFT$ and $\noiseAudioSTFT$ denote the spectrograms of the ground-truth
foreground signal and background noise, respectively. 
The first term penalizes the discrepancy between estimated noise and the ground-truth noise,
while the second term accounts for the estimation of foreground signal.
These two terms are balanced by the scalar $\beta$ ($\beta=1.0$ in our experiments).

\vspace{-0.1mm}
\paraspace
\paragraph{Natural emergence of silent intervals.}
While producing plausible denoising results (see \secref{ablation}), the end-to-end training process has no supervision on silent interval detection:
the loss function~\eq{loss} only accounts for the recoveries of noise and clean speech signal.
But somewhat surprisingly, the ability of detecting silent intervals automatically emerges as the output 
of the first network component (see \figref{results_pipeline}-d as an example, which visualizes $\noiseProfileAudioSTFT$). In other words, the network automatically learns
to detect silent intervals for speech denoising without this supervision.

\vspace{-0.525mm}
\secprespace
\subsection{Silent interval supervision}\label{sec:supervision}
\secspace
As the model is learning to detect silent intervals on its own, we are able to
directly supervise silent interval detection to further improve the denoising
quality. Our first attempt was to add a term in~\eq{loss} that penalizes the
discrepancy between detected silent intervals and their ground truth.  But our
experiments show that this is not effective (see \secref{ablation}).
Instead, we train our network in two sequential steps.

First, we train the silent interval detection component through the following 
loss function:
\begin{equation}\label{eq:si}
    \lossBCE = \mathbb{E}_{\noisyAudio \sim p(\noisyAudio)} \Big[ \ell_\text{BCE} \big(\bm{m}(\noisyAudio), \silentIntervalMask \big) \Big],
\end{equation}
where $\ell_\text{BCE}(\cdot,\cdot)$ is the binary cross entropy loss,
$\bm{m}(\noisyAudio)$ is the mask resulted from silent interval detection component,
and $\silentIntervalMask$ is the ground-truth label of each signal sample being silent or 
not---the way of constructing $\silentIntervalMask$ and the training dataset will be described in \secref{setup}.

Next, we train the noise estimation and removal components through the loss
function~\eq{loss}.  This step starts by neglecting the silent
detection component.  
In the loss function~\eq{loss}, 
instead of using $\noiseProfileAudioSTFT$, the noise
spectrogram exposed by the estimated silent intervals, 
we use the noise spectrogram exposed by the ground-truth silent intervals (i.e., the STFT of $\noisyAudio\odot\silentIntervalMask$).
After training using such a loss function, we fine-tune the network components
by incorporating the already trained silent interval detection component.
With the silent interval detection component fixed, this fine-tuning step optimizes
the original loss function~\eq{loss} and thereby updates the weights
of the noise estimation and removal components.

%% file: 4_Experiments.tex
\vspace{-0.525mm}
\secprespace
\section{Experiments}\label{sec:exp}
\secspace

This section presents the major evaluations of our method, comparisons to
several baselines and prior works, and ablation studies. We also refer the reader to the supplementary materials
(including {a supplemental document and audio effects organized on an off-line webpage}) 
for the full description of our network structure, 
implementation details, additional evaluations, as well as audio examples. 

\vspace{-0.525mm}
\secprespace
\subsection{Experiment setup}\label{sec:setup}
\secspace

\paragraph{Dataset construction.}
To construct training and testing data, we leverage publicly available audio datasets.
We obtain clean speech signals using \AVSP~\cite{Ephrat_2018}, 
from which we randomly choose 2448 videos ($4.5$ hours of total length) and extract their speech audio channels.
Among them, we use 2214 videos for training and 234 videos for testing, so the
training and testing speeches are fully separate. 
All these speech videos are in English, selected on purpose: 
as we show in {supplementary materials},
our model trained on this dataset can readily denoise speeches in other languages.

\begin{figure}[t]
    \centering
    \includegraphics[clip, width=0.92\linewidth]{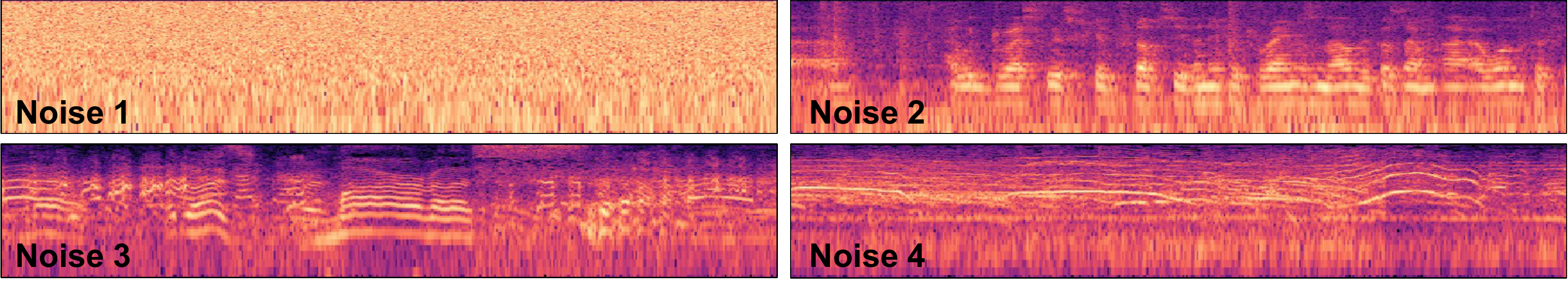}
    \vspace{-2mm}
    \caption{{\bf Noise gallery.} We show four examples of noise from the noise
    datasets. Noise 1) is a stationary (white) noise, and the other three are not. 
    Noise 2) is a monologue in a meeting. Noise 3)
    is party noise from people speaking and laughing with background noise. 
    Noise 4) is street noise from people shouting and screaming
    with additional traffic noise such as vehicles driving and honking.
    }
    \label{fig:noise_examples}
    \vspace{-1mm}
\end{figure}
We use two datasets, \DEM~\cite{thiemann:hal-00796707} and Google's \AS~\cite{audioset},
as background noise. Both consist of 
environmental noise, transportation noise, music, and many other types of noises. 
\DEM has been used in previous 
denoising works (e.g.,~\cite{Pascual2017, Germain2019,
Valentini-Botinhao+2016}). Yet \AS is much larger and more diverse than \DEM, 
thus more challenging when used as noise. Figure~\ref{fig:noise_examples}
shows some noise examples.  Our evaluations are conducted on both datasets,
separately.

Due to the linearity of acoustic wave propagation, we can superimpose clean
speech signals with noise to synthesize noisy input signals (similar to previous works~\cite{Pascual2017, Germain2019, Valentini-Botinhao+2016}).
When synthesizing a noisy input signal, we randomly choose a signal-to-noise ratio
(SNR) from seven discrete values: -10dB, -7dB, -3dB, 0dB, 3dB, 7dB, and
10dB; and by mixing the foreground speech with properly scaled noise, we produce a
noisy signal with the chosen SNR.
For example, a -10dB SNR means that the power of noise is ten times the speech
(see \figref{snr_vary_signals} in appendix).
The SNR range in our evaluations (i.e., [-10dB, 10dB]) is significantly
larger than those tested in previous works.


To supervise our silent interval detection (recall \secref{supervision}), we need ground-truth labels of silent intervals.
To this end, we divide each clean speech signal into time segments, each of which lasts $\nicefrac{1}{30}$ seconds.
We label a time segment as silent when the total acoustic energy in that segment is below a threshold.
Since the speech is clean, this automatic labeling process is robust.

\textit{Remarks on creating our own datasets.}
Unlike many previous models, which are trained using existing datasets such
as Valentini's VoiceBank-DEMAND~\cite{Valentini-Botinhao+2016},
we choose to create our own datasets because of two reasons.
First, Valentini's dataset has a noise SNR level in [0dB, 15dB], 
much narrower than what we encounter in real-world recordings.
Secondly, although Valentini's dataset provides
several kinds of environmental noise, it lacks the richness of other types of
structured noise such as music, making it less ideal for denoising
real-world recordings (see discussion in \secref{rwdata}).

\paraspace
\paragraph{Method comparison.}
We compare our method with several existing methods that are also designed for speech denoising,
including both the classic approaches and recently proposed learning-based methods.
We refer to these methods as follows:
\textbf{i)} \texttt{Ours}, our model trained with silent interval supervision (recall \secref{supervision});
\textbf{ii)} \texttt{Baseline-thres}, a baseline method that uses acoustic
energy threshold to label silent intervals (the same as our automatic labeling
approach in~\secref{setup} but applied on noisy input signals), and then uses
our trained noise estimation and removal networks for speech denoising.
\textbf{iii)} \GTSI, another reference method that uses 
our trained noise estimation and removal networks, but hypothetically uses the ground-truth silent intervals;
\textbf{iv)} \texttt{Spectral Gating}, the classic speech denoising algorithm based on spectral subtraction~\cite{sg};
\textbf{v)} \texttt{Adobe Audition}~\cite{audition}, one of the most widely used professional audio processing software,
and we use its machine-learning-based noise reduction feature, provided in the latest Adobe Audition CC 2020, with default parameters 
to batch process all our test data;
\textbf{vi)} \texttt{SEGAN}~\cite{Pascual2017},
one of the state-of-the-art audio-only speech enhancement methods based on generative adversarial networks.
\textbf{vii)} \texttt{DFL}~\cite{Germain2019}, a recently proposed speech
denoising method based on a loss function over deep network
features;~\footnote{This recent method is designed for high-noise-level input, trained in an end-to-end fashion,
and as their paper states, is ``particularly pronounced for the hardest data
with the most intrusive background noise''.}
\textbf{viii)} \texttt{VSE}~\cite{Gabbay2017}, a learning-based method that takes both 
video and audio as input, and leverages both audio signal and mouth motions (from video footage) for speech denoising.
{We could not compare with another audiovisual method~\cite{Ephrat_2018}
because no source code or executable is made publicly available.}

For fair comparisons, we train all the methods (except \texttt{Spectral Gating} which is not learning-based 
and \texttt{Adobe Audition} which is commercially shipped as a black box)
using the same datasets. For \texttt{SEGAN}, \texttt{DFL}, and \texttt{VSE},
we use their source codes published by the authors.
The audiovisual denoising method \texttt{VSE} also requires video footage,
which is available in \AVSP.

\secprespace
\subsection{Evaluation on speech denoising}\label{sec:eval_denoise}
\secspace

\paragraph{Metrics.} 
Due to the perceptual nature of audio processing tasks, there is no widely
accepted single metric for quantitative evaluation and comparisons.
We therefore evaluate our method under six different metrics, all of which have been frequently used
for evaluating audio processing quality.
Namely, these metrics are:
\textbf{i)} Perceptual Evaluation of Speech Quality (PESQ)~\cite{pesq}, 
\textbf{ii)} Segmental Signal-to-Noise Ratio (SSNR)~\cite{ssnr},
\textbf{iii)} Short-Time Objective Intelligibility (STOI)~\cite{stoi}, 
\textbf{iv)} Mean opinion score (MOS) predictor of signal distortion (CSIG)~\cite{metrics}, 
\textbf{v)} MOS predictor of background-noise intrusiveness (CBAK)~\cite{metrics}, 
and \textbf{vi)} MOS predictor of overall signal quality (COVL)~\cite{metrics}.

\begin{figure}[t]
\vspace{-0mm}
\includegraphics[width=\linewidth]{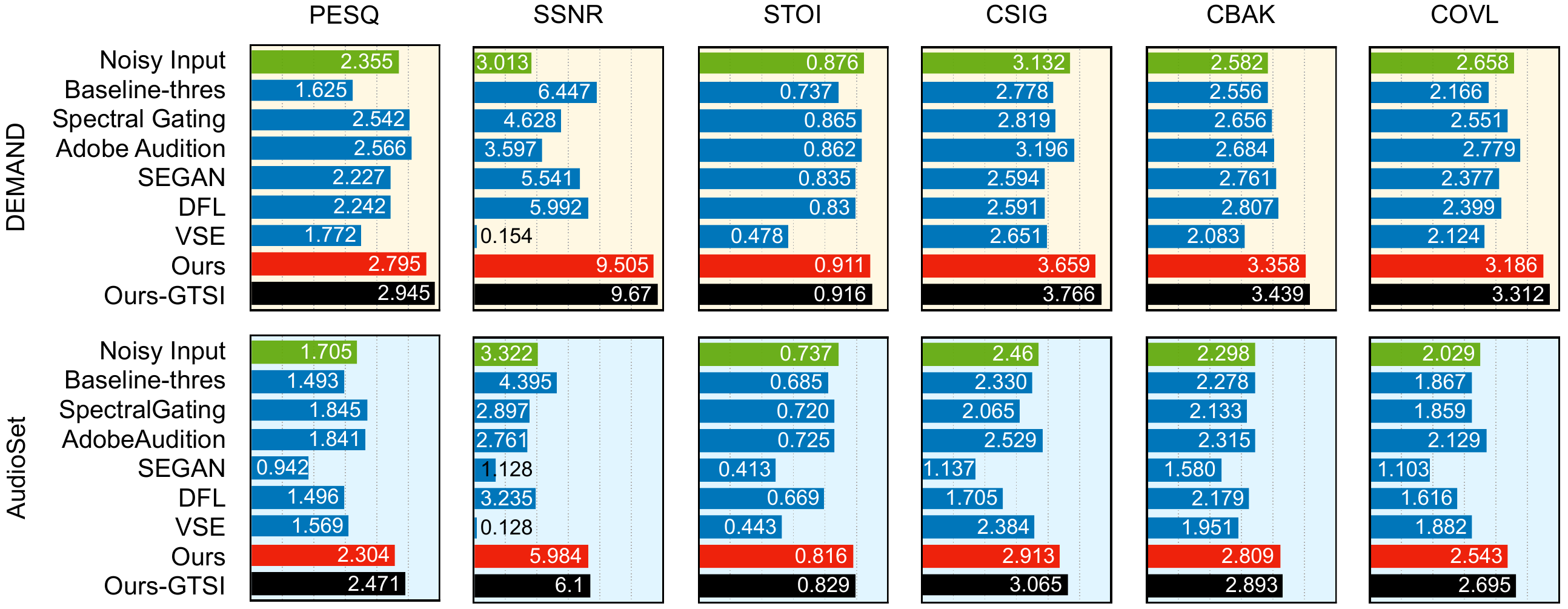}
\vspace{-6mm}
\caption{\textbf{Quantitative comparisons.}
We measure denoising quality under six metrics (corresponding to columns).
The comparisons are conducted using noise from \DEM and \AS separately.
\GTSI (in black) uses ground-truth silent intervals. Although not a practical
approach, it serves as an upper-bound reference of all methods. 
Meanwhile, the green bar in each plot indicates the metric score of the noisy
input without any processing.
\label{fig:quan-denoise}}
\end{figure}

\begin{figure}[t]
\vspace{-2.5mm}
\includegraphics[width=0.90\linewidth]{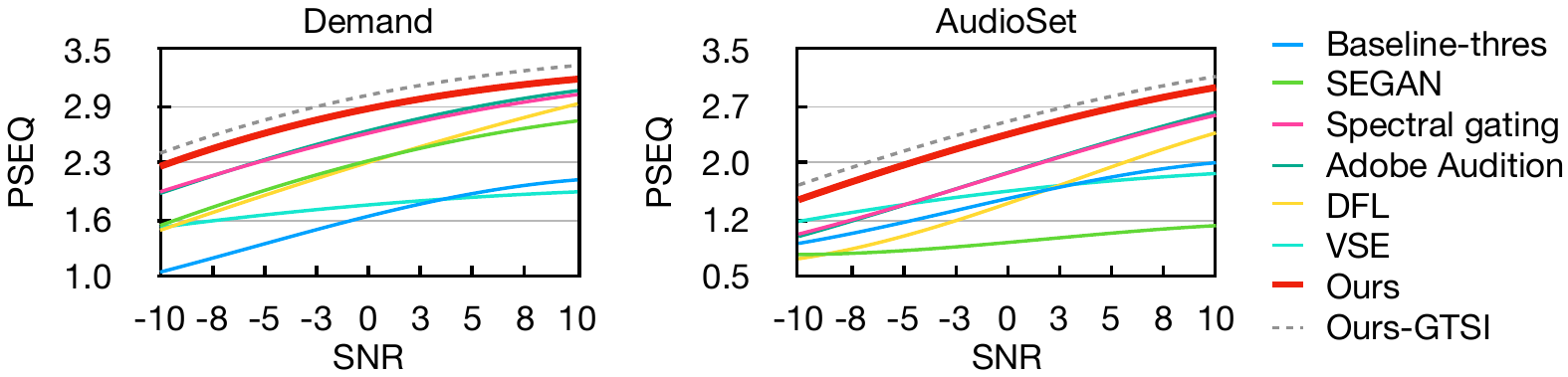}
\vspace{-3mm}
\caption{\textbf{Denoise quality w.r.t. input SNRs.} Denoise results
measured in PESQ for each method w.r.t different input SNRs.
Results measured in other metrics are shown in \figref{ad-comp-all} in Appendix.}
\label{fig:ad-comp}
\end{figure}

\paraspace
\paragraph{Results.}
We train two separate models using \DEM and \AS noise datasets respectively,
and compare them with other models trained with the same datasets.
We evaluate the average metric values and report them in \figref{quan-denoise}.
Under all metrics, our method consistently outperforms others.

We breakdown the performance of each method with respect to SNR levels from -10dB to 10dB on both noise datasets. The results are reported in \figref{ad-comp} for PESQ (see \figref{ad-comp-all} in the appendix for all metrics).
In the previous works that we compare to, no results under those low SNR levels (at $<0$ dBs) are reported. 
Nevertheless, across all input SNR levels, our method performs the best, showing that our approach is fairly robust to both light and extreme noise.

From \figref{ad-comp}, it is worth noting that \GTSI method performs even
better. Recall that this is our model but provided with ground-truth silent
intervals. While not practical (due to the need of ground-truth silent
intervals), \GTSI confirms the importance of silent intervals for denoising:
a high-quality silent interval detection helps to improve speech denoising
quality.

\secprespace
\subsection{Evaluation on silent interval detection}\label{sec:sid}
\secspace

Due to the importance of silent intervals for speech denoising,
we also evaluate the quality of our silent interval detection, in comparison to two 
alternatives, the baseline \texttt{Baseline-thres} and a Voice
Activity Detector (\texttt{VAD})~\cite{vad}. The former is described above, while the latter
classifies each time window of an audio signal as having human voice or not~\cite{293048, leb}.
We use an off-the-shelf VAD~\cite{webrtcvad}, which is developed by Google's WebRTC project and reported as one
of the best available. 
{Typically, VAD is designed to work with low-noise signals. Its inclusion here
here is merely to provide an alternative approach that can detect silent
intervals in more ideal situations.}

We evaluate these methods using four standard statistic metrics: the precision, recall,
F1 score, and accuracy. We follow the standard definitions of these metrics, which are summarized in \appref{sid_metrics}. 
These metrics are based on the definition of positive/negative conditions.
Here, the positive condition indicates a time segment being labeled as a silent segment, and
the negative condition indicates a non-silent label.
Thus, the higher the metric values are, the better the detection approach.

\begin{table}[t]
\caption{\textbf{Results of silent interval detection.} The metrics are measured using our test signals that have
SNRs from -10dB to 10dB. Definitions of these metrics are summarized in \appref{sid_metrics}.
\label{tab:sid}}
\vspace{-3mm}
\centering
\begin{tabular}{llcccccc} \toprule
                      Noise Dataset & Method                  & Precision & Recall & F1       & Accuracy \\ \midrule
    \multirow{3}{*}{DEMAND}   & \texttt{Baseline-thres} & 0.533 & 0.718 & 0.612          & 0.706 \\ 
                              & \texttt{VAD}            & 0.797 & 0.432 & 0.558 & 0.783 \\
                              & \texttt{Ours}           & \textbf{0.876} & \textbf{0.866} & \textbf{0.869} & \textbf{0.918} \\ \midrule
    \multirow{3}{*}{Audioset} & \texttt{Baseline-thres} & 0.536 & 0.731 & 0.618          & 0.708 \\ 
                              & \texttt{VAD}            & 0.736 & 0.227 & 0.338 & 0.728 \\
                              & \texttt{Ours}           & \textbf{0.794} & \textbf{0.822} & \textbf{0.807} & \textbf{0.873} \\ \bottomrule
\end{tabular}
\vspace{-1mm}
\end{table}

\tabref{sid} shows that, under all metrics, our method is consistently better than the alternatives.
Between \texttt{VAD} and \texttt{Baseline-thres}, \texttt{VAD} has higher
precision and lower recall, meaning that \texttt{VAD} is overly conservative
and \texttt{Baseline-thres} is overly aggressive when detecting silent
intervals (see \figref{sid} in \appref{sid_result}).
Our method reaches better balance and thus detects silent intervals more accurately.

\secprespace
\subsection{Ablation studies}\label{sec:ablation}
\secspace

\begin{table}[t]
\caption{\textbf{Ablation studies.} We alter network components and training loss, and evaluate the denoising quality under various metrics.
Our proposed approach performs the best.
\label{tab:ablation_all}}
\vspace{-3mm}
\begin{tabular}{llcccccc} \toprule
            Noise Dataset & Method & PESQ  & SSNR  & STOI  & CSIG  & CBAK  & COVL  \\ \midrule
\multirow{6}{*}{DEMAND}   
                          & \texttt{Ours w/o SID comp} & 2.689 & 9.080 & 0.904 & 3.615 & 3.285 & 3.112 \\
                          & \texttt{Ours w/o NR comp}  & 2.476 & 0.234 & 0.747 & 3.015 & 2.410 & 2.637 \\
                          & \texttt{Ours w/o SID loss} & 2.794 & 6.478 & 0.903 & 3.466 & 3.147 & 3.079 \\
                          & \texttt{Ours w/o NE loss}  & 2.601 & 9.070 & 0.896 & 3.531 & 3.237 & 3.027 \\
                          & \texttt{Ours Joint loss} & 2.774 & 6.042 & 0.895 & 3.453 & 3.121 & 3.068 \\
                          & \texttt{Ours} & \textbf{2.795} & \textbf{9.505} & \textbf{0.911} & \textbf{3.659} & \textbf{3.358} & \textbf{3.186} \\ \midrule
\multirow{6}{*}{Audioset} 
                          & \texttt{Ours w/o SID comp} & 2.190 & 5.574 & 0.802 & 2.851 & 2.719 & 2.454 \\
                          & \texttt{Ours w/o NR comp}  & 1.803 & 0.191 & 0.623 & 2.301 & 2.070 & 1.977 \\
                          & \texttt{Ours w/o SID loss} & \textbf{2.325} & 4.957 & 0.814 & 2.814 & 2.746 & 2.503 \\
                          & \texttt{Ours w/o NE loss}  & 2.061 & 5.690 & 0.789 & 2.766 & 2.671 & 2.362 \\
                          & \texttt{Ours Joint loss} & 2.305 & 4.612 & 0.807 & 2.774 & 2.721 & 2.474 \\
                          & \texttt{Ours} & 2.304 & \textbf{5.984} & \textbf{0.816} & \textbf{2.913} & \textbf{2.809} & \textbf{2.543} \\ \bottomrule
\end{tabular}
\vspace{0.5mm}
\end{table}

In addition, we perform a series of ablation studies to understand the efficacy
of individual network components and loss terms (see \appref{aa_details} for more details). 
In \tabref{ablation_all}, ``\texttt{Ours w/o SID loss}'' refers to the training method presented in \secref{training}
(i.e., without silent interval supervision).
``\texttt{Ours Joint loss}'' refers to the end-to-end training approach that optimizes the loss function~\eq{loss} with the additional
term~\eq{si}. And ``\texttt{Ours w/o NE loss}'' uses our two-step training (in \secref{supervision}) but without the loss term on
noise estimation---that is, without the first term in \eq{loss}. In comparison to these alternative training approaches, our two-step training with silent interval supervision
(referred to as ``\texttt{Ours}'') performs the best. 
We also note that 
``\texttt{Ours w/o SID loss}''---i.e., without supervision on silent interval 
detection---already outperforms the methods we compared to in \figref{quan-denoise},
and ``\texttt{Ours}'' further improves the denoising quality.
This shows the efficacy of our proposed training approach.

We also experimented with two variants of our network structure. The first one,
referred to as ``\texttt{Ours w/o SID comp}'', turns off silent interval detection: the silent interval detection 
component always outputs a vector with all zeros. 
The second, referred as ``\texttt{Ours w/o NR comp}'', uses a simple spectral subtraction 
to replace our noise removal component. \tabref{ablation_all} shows that, under all the tested metrics, both variants perform
worse than our method, suggesting our proposed network structure is effective.


Furthermore, we studied to what extent the accuracy of silent interval detection affects the speech denoising quality.
We show that as the silent interval detection becomes less accurate, the denoising quality degrades. 
Presented in details in \appref{sid_robustness}, these experiments reinforce our intuition that silent intervals
are instructive for speech denoising tasks.

\secprespace
\subsection{Comparison with state-of-the-art benchmark}\label{sec:benchmark}
\secspace

Many state-of-the-art denoising methods, including MMSE-GAN~\cite{tfgan},
Metric-GAN~\cite{fu2019metricgan}, SDR-PESQ~\cite{kim2019endtoend},
T-GSA~\cite{tgsa}, Self-adaptation DNN~\cite{koizumi2020speech}, and
RDL-Net~\cite{nikzad2020deep}, are all evaluated
on Valentini's VoiceBank-DEMAND~\cite{Valentini-Botinhao+2016}.
We therefore compare ours with those methods on the same dataset. 
We note that DEMAND consists of audios with SNR in [0dB, 15dB]. Its SNR range is 
much narrower than what our method (and our training datasets) aims for (e.g., input signals with -10dB SNR).
Nevertheless, trained and tested under the same setting, our method is highly competitive to
the best of those methods under every metric, 
as shown in \tabref{benchmark}. The metric scores therein for other methods are
numbers reported in their original papers.

\begin{table}[t]
\begin{tabular}{lccccc} \toprule
Method & PESQ  & CSIG  & CBAK  & COVL  & STOI\\ \midrule
\texttt{Noisy Input} & 1.97 & 3.35 & 2.44 & 2.63 & 0.91 \\
\texttt{WaveNet}~\cite{Oord2016aiw} & -- & 3.62 & 3.24 & 2.98 & -- \\
\texttt{SEGAN}~\cite{Pascual2017} & 2.16 & 3.48 & 2.94 & 2.80 & 0.93 \\
\texttt{DFL}~\cite{Germain2019} & 2.51 & 3.79 & 3.27 & 3.14 & -- \\
\texttt{MMSE-GAN}~\cite{tfgan} & 2.53 & 3.80 & 3.12 & 3.14 & 0.93 \\
\texttt{MetricGAN}~\cite{fu2019metricgan} & 2.86 & 3.99 & 3.18 & 3.42 & -- \\
\texttt{SDR-PESQ}~\cite{kim2019endtoend} & 3.01 & 4.09 & 3.54 & 3.55 & -- \\
\texttt{T-GSA}~\cite{tgsa} & 3.06 & 4.18 & {3.59} & 3.62 & -- \\
\texttt{Self-adapt.~DNN}~\cite{koizumi2020speech} & 2.99 & 4.15 & 3.42 & 3.57 & -- \\
\texttt{RDL-Net}~\cite{nikzad2020deep} & 3.02 & {4.38} & 3.43 & {3.72} & 0.94 \\
\texttt{Ours} & {3.16} & 3.96 & 3.54 & 3.53 & {0.98} \\ \bottomrule
\end{tabular}
\caption{Comparisons on VoiceBank-DEMAND corpus.}
\label{tab:benchmark}
\end{table}

\secprespace
\subsection{Tests on real-world data}\label{sec:rwdata}
\secspace
We also test our method against real-world data.  
Quantitative evaluation on real-world data, however, is not easy because the evaluation 
of nearly all metrics requires the corresponding ground-truth clean signal, which is not
available in real-world scenario.  Instead, we collected a good number of
real-world audios, either by recording in daily environments or
by downloading online (e.g., from YouTube).  These real-world audios cover diverse
scenarios: in a driving car, a caf\'{e}, a park, on the street, in multiple
languages (Chinese, Japanese, Korean, German, French, etc.), with
different genders and accents, and even with singing songs. None of these
recordings is cherry picked. We refer the reader to our project website for the
denoising results of all the collected real-world recordings, and for the comparison of 
our method with other state-of-the-art methods under real-world settings.

Furthermore, we use real-world data to test our model trained with different datasets, including
our own dataset (recall \secref{setup}) and the existing 
DEMAND~\cite{Valentini-Botinhao+2016}. We show that the network model trained by our own dataset leads to 
much better noise reduction (see details in \appref{rw}).
This suggests that our dataset allows the denoising model to better generalize to 
many real-world scenarios.

%% file: 5_Conclusion.tex
\secprespace
\section{Conclusion}
\secspace
Speech denoising has been a long-standing challenge. We present a new network structure
that leverages the abundance of silent intervals in speech. 
Even without silent interval supervision, 
our network is able to denoise speech signals plausibly,
and meanwhile, the ability to detect silent intervals automatically emerges.
We reinforce this ability.
Our explicit supervision on silent intervals enables the network to
detect them more accurately, thereby further improving the
performance of speech denoising.
As a result, under a variety of denoising metrics, our method consistently outperforms 
several state-of-the-art audio denoising models.

%

%
%

%% file: 6_Broader.tex
\section*{Broader Impact}
\secspace
High-quality speech denoising is desired in a myriad of applications: human-robot
interaction, cellular communications, hearing aids, teleconferencing, music
recording, filmmaking, news reporting, and surveillance systems to name a few.
Therefore, we expect our proposed denoising method---be it a system used in practice or
a foundation for future technology---to find impact in these applications. 

In our experiments,
we train our model using English speech only, to demonstrate 
its generalization property---the ability of denoising spoken languages beyond English.
Our demonstration of denoising Japanese, Chinese, and Korean speeches 
is intentional: they are linguistically and phonologically distant from
English (in contrast to other English ``siblings'' such as {German} and {Dutch}).
Still, our model may bias in favour of spoken languages and cultures
that are closer to English or that have frequent pauses to reveal silent intervals.
Deeper understanding of this potential bias
requires future studies in tandem with linguistic and sociocultural insights.

Lastly, it is natural to extend our model for denoising 
audio signals in general or even signals beyond audio (such as Gravitational wave denoising~\cite{wei2020gravitational}).
If successful, our model can bring in even broader impacts.
Pursuing this extension, however, requires a judicious definition of ``silent intervals''.
After all, the notion of ``noise'' in a general context of signal processing 
depends on specific applications: noise in one application may be another's signals.
To train a neural network that exploits a general notion of silent intervals,
prudence must be taken to avoid biasing toward certain types of noise.


%% file: app.tex
\setcounter{figure}{0}    
\renewcommand{\thefigure}{S\arabic{figure}}
\setcounter{table}{0}
\renewcommand{\thetable}{S\arabic{table}}
\setcounter{equation}{0}
\renewcommand{\theequation}{S\arabic{equation}}

\vspace{10mm}
\begin{center}
\Large
\textbf{Supplementary Document}\\ 
\smallskip
\textbf{Listening to Sounds of Silence for Speech Denoising}
\medskip
\end{center}


\vspace{10mm}
\section{Network Structure and Training Details}\label{sec:network_details}
We now present the details of our network structure and training configurations.

\textbf{The silent interval detection component} of our model is composed of 2D
convolutional layers, a bidirectional LSTM, and two FC layers. The parameters of
the convolutional layers are shown in ~\tabref{sid_param}. Each convolutional
layer is followed by a batch normalization layer with a ReLU activation
function. The hidden size of bidirectional LSTM is $100$. The two FC layers,
interleaved with a ReLU activation function, have hidden size of $100$ and $1$, respectively. 

\begin{table}[h]
\centering
\caption{Convolutional layers in the silent interval detection component.
\label{tab:sid_param}}
\vspace{-2mm}
\resizebox{\linewidth}{!}{
\begin{tabular}{lcccccccccccc} \toprule
            & conv1 & conv2 & conv3 & conv4 & conv5 & conv6 & conv7 & conv8 & conv9 & conv10 & conv11 & conv12 \\ \midrule
Num Filters &  48   & 48 & 48 & 48 & 48 & 48 & 48 & 48 & 48 & 48 & 48 & 8 \\
Filter Size & (1,7) & (7,1) & (5,5) & (5,5) & (5,5) & (5,5) & (5,5) & (5,5) & (5,5) & (5,5) & (5,5) & (1,1)\\
Dilation    & (1,1) & (1,1) & (1,1) & (2,1) & (4,1) & (8,1) & (16,1) & (32,1) & (1,1) & (2,2) & (4,4) & (1,1)\\
Stride      &  1   & 1 & 1 & 1 & 1 & 1 & 1 & 1 & 1 & 1 & 1 & 1 \\ \bottomrule
\end{tabular}
}
\end{table}


\textbf{The noise estimation component} of our model is fully convolutional,
consisting of two encoders and one decoder.
The two encoders process the noisy signal and the incomplete noise profile, respectively; 
they have the same architecture (shown in \tabref{ne_param}) but different weights. 
The two feature maps resulted from the two encoders are concatenated in a
channel-wise manner before feeding into the decoder.
In \tabref{ne_param}, every layer, except the last one, is followed by a batch
normalization layer together with a ReLU activation function. In addition, there is skip
connections between the $2$nd and $14$-th layer and between the $4$-th and $12$-th layer.

\begin{table}[h]
\centering
\caption{Architecture of noise estimation component. `C' indicates a convolutional layer, 
    and `TC' indicates a transposed convolutional layer.
\label{tab:ne_param}}
\vspace{-2mm}
\resizebox{\linewidth}{!}{
\begin{tabular}{l ccccccccccccccccc}\toprule
    & \multicolumn{3}{ c }{\textbf{Encoder}}& \phantom{a}&\multicolumn{13}{c}{\textbf{Decoder}}\\ 
            \cmidrule{2-4} \cmidrule{6-18}
ID          &  1    &  2    &  3    &&  4    &  5    &  6    &  7    &  8    &  9    & 10    &  11   &  12   &  13   &  14   &  15   &  16   \\\midrule
Layer Type  &  \texttt{C}    &  \texttt{C}    &  \texttt{C}    &&  \texttt{C}    &  \texttt{C}    &  \texttt{C}     &  \texttt{C}     &  \texttt{C}    &  \texttt{C}    &  \texttt{C}    &  \texttt{C}    &  \texttt{TC}   &  \texttt{C}  &  \texttt{TC}   &  \texttt{C}  &  \texttt{C} \\ 
Num Filters &  64   & 128   & 128   && 256   &   256 &   256 &   256 &   256 &   256 &   256 &   256 &   128 &   128 &   64  &   64  &   2 \\
Filter Size &   5   &     5 &     5 &&   3   &   3   &   3   &   3   &   3   &   3   &   3   &   3   &   3   &   3   &   3   &   3   &   3  \\
Dilation    &   1   &   1   &   1   &&   1   &   1   &   2   &   4   &   8   &   16  &   1   &   1   &   1   &   1   &   1   &   1   &   1  \\
Stride      &  1    &   2   &  1    &&  2    &     1 &     1 &     1 &     1 &     1 &     1 &     1 &     2 &     1 &     2 &     1 &     1  \\\bottomrule
\end{tabular}
}
\end{table}

\textbf{The noise removal component} of our model is composed of two 2D
convolutional encoders, a bidirectional LSTM, and three FC layers. 
The two convolutional encoders take as input the input audio spectrogram $\noisyAudioSTFT$ 
and the estimated full noise spectrogram $\nRecoverModule(\noisyAudioSTFT, \noiseProfileAudioSTFT)$,
respectively. The first encoder has the network architecture listed in \tabref{nr_param},
and the second has the same architecture but with half of the number of filters 
at each convolutional layer.
Moreover, the bidirectional LSTM has the hidden size of 200,
and the three FC layers have the hidden size of 600, 600, and $2F$, respectively, 
where $F$ is the number of frequency bins in the spectrogram.
In terms of the activation function, ReLU is used after each layer except the last layer, which uses sigmoid.


\begin{table}[h]
\caption{Convolutional encoder for the noise removal component of our model. Each convolutional layer is followed by a batch normalization layer with ReLU as the activation function.}
\label{tab:nr_param}
\vspace{-2mm}
\centering
\resizebox{\linewidth}{!}{
\begin{tabular}{lccccccccccccccc}\toprule
            & C1 & C2 & C3 & C4 & C5 & C6 & C7 & C8 & C9 & C10 & C11 & C12 & C13 & C14 & C15 \\ \midrule
Num Filters &  96   & 96 & 96 & 96 & 96 & 96 & 96 & 96 & 96 & 96 & 96 & 96 & 96 & 96 & 8 \\
Filter Size & (1,7) & (7,1) & (5,5) & (5,5) & (5,5) & (5,5) & (5,5) & (5,5) & (5,5) & (5,5) & (5,5) & (5,5) & (5,5) & (5,5) & (1,1)\\
Dilation    & (1,1) & (1,1) & (1,1) & (2,1) & (4,1) & (8,1) & (16,1) & (32,1) & (1,1) & (2,2) & (4,4) & (8,8) & (16,16) & (32,32) & (1,1)\\
Stride      &  1   & 1 & 1 & 1 & 1 & 1 & 1 & 1 & 1 & 1 & 1 & 1 & 1 & 1 & 1
\\\bottomrule
\end{tabular}
}
\end{table}

\paragraph{Training details.}
We use PyTorch platform to implement our speech denoising model,
which is then trained with the Adam optimizer.
In our end-to-end training without silent interval supervision (referred to as ``\texttt{Ours w/o SID loss}'' in \secref{exp}; also recall \secref{training}),
we run the Adam optimizer for 50 epochs with a batch size of 20 and 
a learning rate of 0.001.
When the silent interval supervision is incorporated (recall \secref{supervision}),
we first train the silent interval detection component with the following setup:
run the Adam optimizer for 100 epochs with a batch size of 15 and a learning rate of 0.001.
Afterwards, we train the noise estimation and removal components using the same setup 
as the end-to-end training of ``\texttt{Ours w/o SID loss}''.


\section{Data Processing Details}



Our model is designed to take as input a mono-channel audio clip of an
arbitrary length. However, when constructing the training dataset, we 
set each audio clip in the training dataset to have the same 2-second length, to 
enable batching at training time.
To this end, we split each original audio clip from \AVSP, \DEM, and \AS into 2-second long clips.
All audio clips are then downsampled at 16kHz before converting into spectrograms using STFT.
To perform STFT, the Fast Fourier Transform (FFT) size is set to 510, the
Hann window size is set to 28ms, and the hop length is set to 11ms.
As a result, each 2-second clip yields a (complex-valued) spectrogram with a resolution 
$256\times178$, where $256$ is the number of frequency bins, and 178 is the temporal resolution.
At inference time, our model can still accept audio clips with arbitrary length.

Both our clean speech dataset and noise datasets are first split into training
and test sets, so that no audio clips in training and testing are from the same original 
audio source---they are fully separate. 



To supervise our silent interval detection, we label the clean audio signals in 
the following way.
We first normalize each audio clip so that its magnitude is in the range [-1,1],
that is, ensuring the largest waveform magnitude at -1 or 1.
Then, the clean audio clip is divided into segments of length $\nicefrac{1}{30}$ seconds. 
We label a time segment as a ``silent'' segment (i.e., label 0) if
its average waveform energy in that segment is below 0.08. Otherwise, it is
labeled as a ``non-silent'' segment (i.e., label 1).


\begin{figure}[t]
    \centering
    \includegraphics[page=1, clip, trim=1.125in 8.875in 1.125in 0.75in, width=0.96\linewidth]{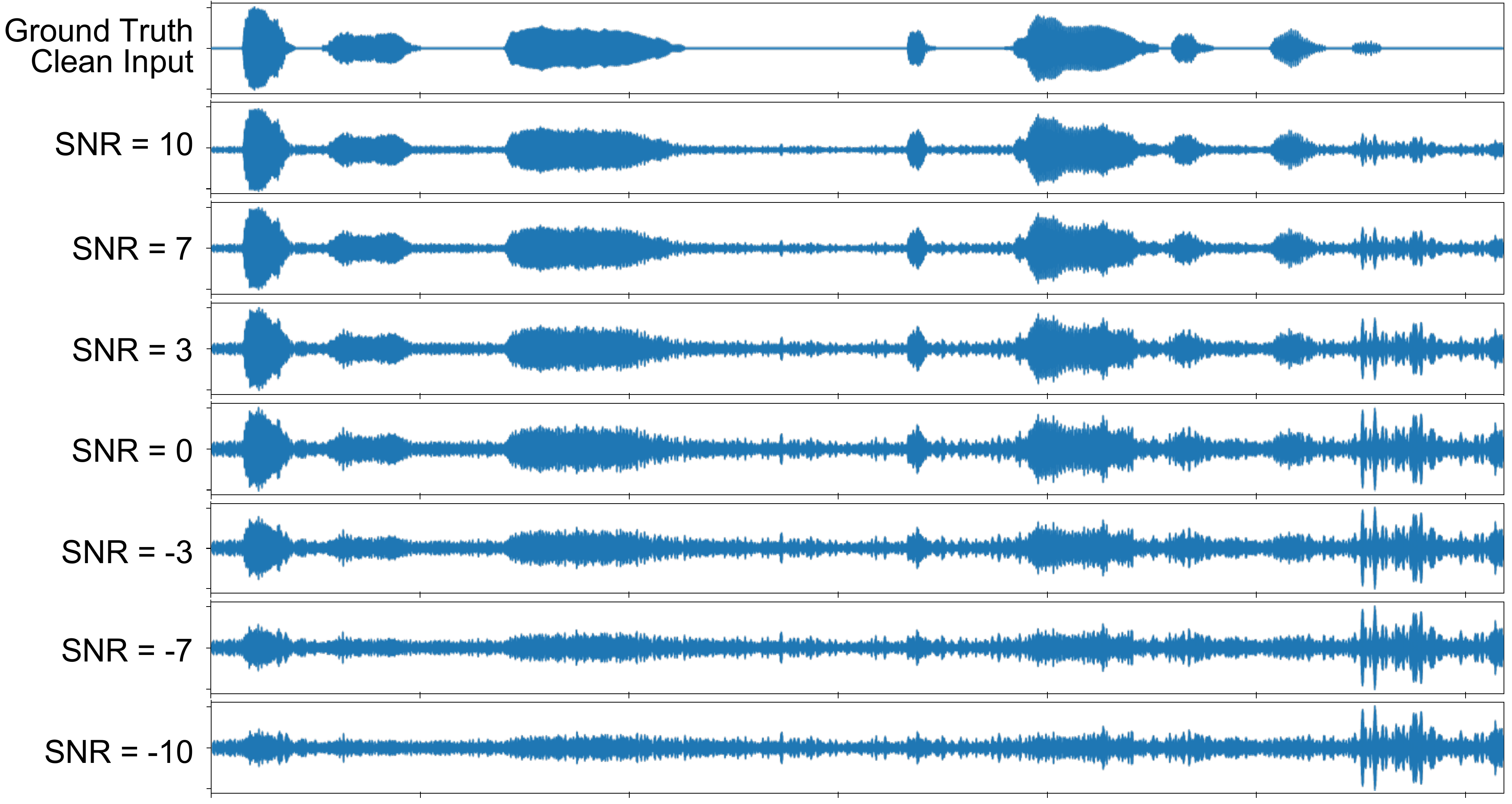}
    \vspace{-3mm}
    \caption{{\bf Constructed noisy audio based on different SNR levels.} The first row shows the waveform of the ground truth clean input.}
    \label{fig:snr_vary_signals}
\end{figure}



\newpage
\begin{figure}[h]
\includegraphics[width=0.8\linewidth]{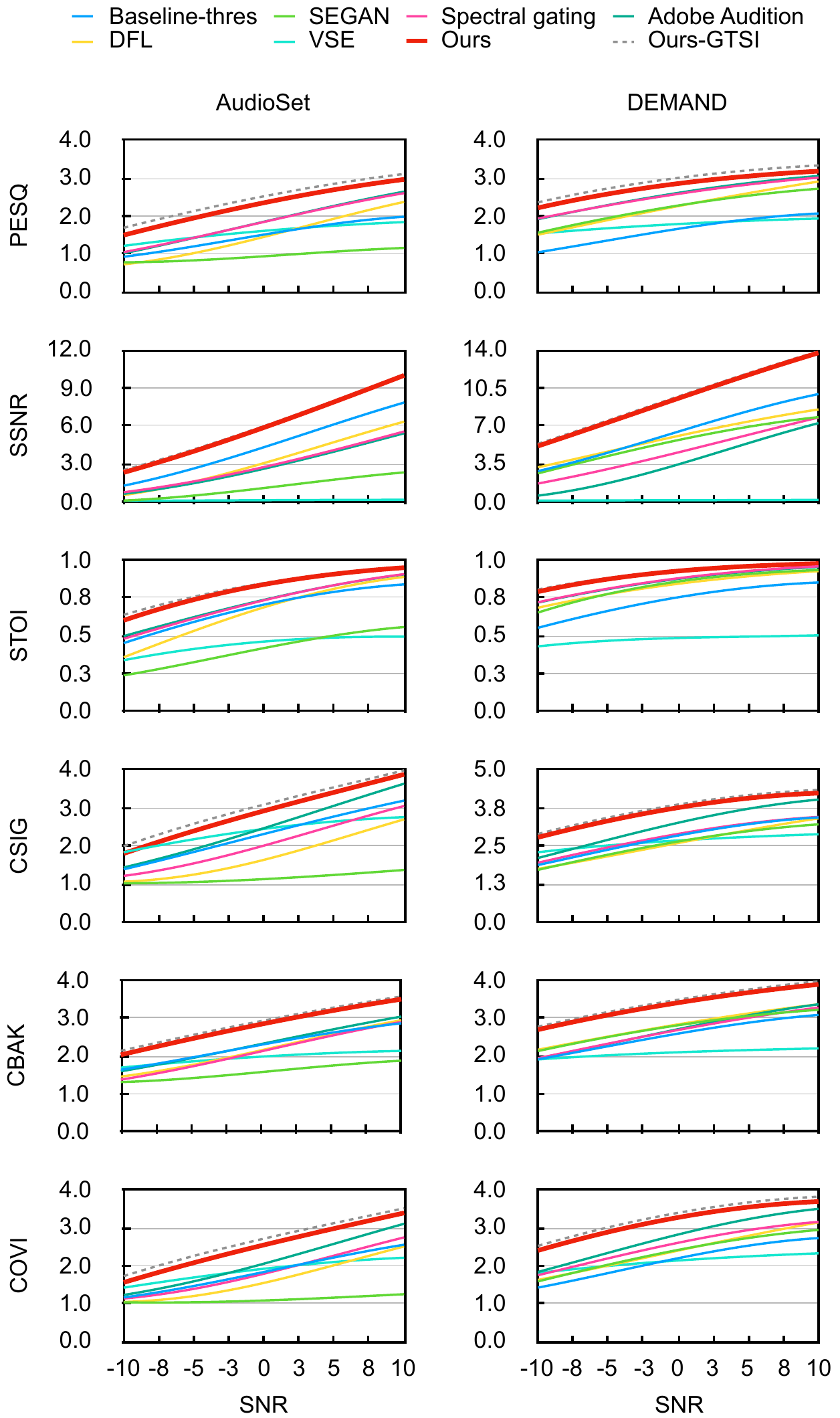}
\vspace{-2mm}
\caption{\textbf{Denoise quality under different input SNRs.} 
Here we expand \figref{ad-comp} of the main text, including 
the evaluations under all six metrics described in \secref{eval_denoise}.}
\label{fig:ad-comp-all}
\end{figure}

\section{Evaluation on Silent Interval Detection}

\subsection{Metrics}\label{sec:sid_metrics}
We now provide the details of the metrics used for evaluating our silent 
interval detection (i.e., results in \tabref{sid} of the main text).
Detecting silent intervals is a binary classification task, one that classifies
every time segment as being silent (i.e., a positive condition) or not (i.e., a negative condition).
Recall that the confusion matrix in a binary classification task is as follows:
\vspace{-2mm}
{\renewcommand{\arraystretch}{2}%
\begin{table}[H]
\caption{Confusion matrix}
\vspace{-2mm}
\centering
\begin{tabular}{@{}cc cc@{}}\toprule
\multicolumn{1}{c}{} &\multicolumn{1}{c}{} &\multicolumn{2}{c}{Actual} \\ 
\cmidrule{3-4}
\multicolumn{1}{c}{} & \multicolumn{1}{c}{} & \multicolumn{1}{c}{Positive} & \multicolumn{1}{c}{Negative} \\ 
\cmidrule{2-4}
\multirow[c]{2}{*}{\rotatebox[origin=tr]{90}{Predicted}}
& Positive  & True Positive ($\textsf{TP}$)  & False Positive ($\textsf{FP}$) \\
& Negative  & False Negative ($\textsf{FN}$) & True Negative ($\textsf{TN}$) \\ 
\bottomrule
\end{tabular}
\end{table}}
\vspace{-3mm}


In our case, we have the following conditions:
\begin{itemize}
    \item A true positive ($\textsf{TP}$) sample is a correctly predicted silent segment.
    \item A true negative ($\textsf{TN}$) sample is a correctly predicted non-silent segment.
    \item A false positive ($\textsf{FP}$) sample is a non-silent segment predicted as silent.
    \item A false negative ($\textsf{FN}$) sample is a silent segment predicted as non-silent.
\end{itemize}
The four metrics used in \tabref{sid} follow the standard definitions in statistics, which
we review here:
\begin{equation} 
    \begin{split}
        \textrm{precision} &= \frac{N_{\textsf{TP}}}{N_{\textsf{TP}} + N_{\textsf{FP}}}, \\
        \textrm{recall} &= \frac{N_{\textsf{TP}}}{N_{\textsf{TP}} + N_{\textsf{FN}}}, \\
        \textrm{F1} &= 2 \cdot \frac{\textrm{precision} \cdot \textrm{\textrm{recall}}}{\textrm{precision} + \textrm{recall}}, \textrm{ and } \\
        \textrm{accuracy}& = \frac{N_{\textsf{TP}} + N_{\textsf{TN}}}{N_{\textsf{TP}} + N_{\textsf{TN}} + N_{\textsf{FP}} + N_{\textsf{FN}}},
    \end{split}
\end{equation}
where $N_{\textsf{TP}}$, $N_{\textsf{TN}}$, $N_{\textsf{FP}}$, and $N_{\textsf{FN}}$ indicate
the numbers of true positive, true negative, false positive, and false negative predictions
among all tests.
Intuitively, \emph{recall} indicates the ability of correctly finding all true silent intervals,
\emph{precision} measures how much proportion of the labeled silent intervals 
are truly silent. \emph{F1} score 
takes both precision and recall into account, and produces their harmonic mean.
And \emph{accuracy} is the ratio of correct predictions among all predictions.


\subsection{An Example of Silent Interval Detection}\label{sec:sid_result}
\begin{figure}[b]
    \centering
    \includegraphics[width=0.98\linewidth]{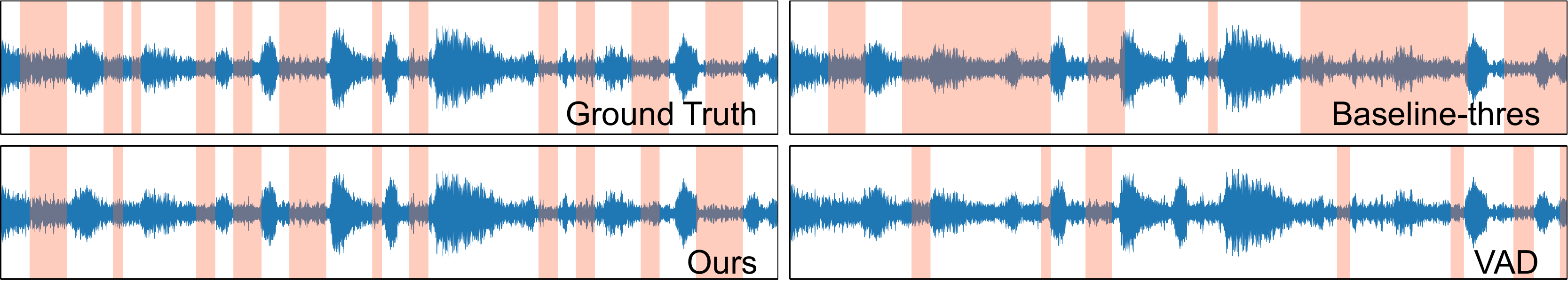}
    \vspace{-2mm}
    \caption{{\bf An example of silent interval detection results.} 
Provided an input signal whose SNR is 0dB (top-left),
we show the silent intervals (in red) detected by three approaches: our method, 
\texttt{Baseline-thres}, and \texttt{VAD}. We also show ground-truth silent intervals
in top-left.
    \label{fig:sid}}
\end{figure}

In \figref{sid}, we present an example of silent interval detection results 
in comparison to two alternative methods.
The two alternatives, described in \secref{sid}, are referred to as 
\texttt{Baseline-thres} and \texttt{VAD}, respectively.
Figure~\ref{fig:sid} echos the quantitative results in \tabref{sid}:
\texttt{VAD} tends to be overly conservative, even in the presence of mild noise;
and many silent intervals are ignored.
On the other hand,
\texttt{Baseline-thres} tends to be too aggressive; it produces many false
intervals.  In contrast, our silent interval detection maintains a better
balance, and thus predicts more accurately.

\section{Ablation Studies and Analysis}

\subsection{Details of Ablation Studies}\label{sec:aa_details}

In \secref{ablation} and \tabref{ablation_all}, the ablation studies are set up in the following 
way.
\begin{itemize}
    \item ``\texttt{Ours}'' refers to our proposed network structure and training method that 
        incorporates silent interval supervision (recall \secref{supervision}).
        Details are described in \appref{network_details}.
    \item ``\texttt{Ours w/o SID loss}'' refers to our proposed network structure
        but optimized by the training method in \secref{training} (i.e. an end-to-end training 
        without silent interval supervision). 
        This ablation study is to confirm that silent interval supervision indeed helps to 
        improve the denoising quality.
    \item ``\texttt{Ours Joint loss}'' refers to our proposed network structure
        optimized by the end-to-end training approach that optimizes 
        the loss function~\eq{loss} with the additional term~\eq{si}. 
        In this end-to-end training, silent interval detection
        is also supervised through the loss function.
        This ablation study is to confirm that our two-step training (\secref{supervision}) is more effective.
    \item ``\texttt{Ours w/o NE loss}'' uses our two-step training (in
        \secref{supervision}) but without the loss term on noise
        estimation---that is, without the first term in \eq{loss}. 
        This ablation study is to examine the necessity of the loss term on noise estimation 
        for better denoising quality.
    \item ``\texttt{Ours w/o SID comp}'' turns off silent interval detection: the
        silent interval detection component always outputs a vector with all
        zeros. As a result, the input noise profile to the noise estimation
        component $\nRecoverModule$ is made precisely the same as the original noisy
        signal. This ablation study is to examine the effect of 
        silent intervals for speech denoising.
    \item ``\texttt{Ours w/o NR comp}'' uses a simple spectral subtraction to
        replace our noise removal component; the other components remain unchanged. 
        This ablation studey is to examine the efficacy of our noise removal component.
\end{itemize}

\subsection{The Influence of Silent Interval Detection on Denoising Quality}
\label{sec:sid_robustness}
A key insight of our neural-network-based denoising model is the leverage of
silent interval distribution over time.  The experiments above have confirmed the efficacy of our
silent interval detection for better speech denoising.
We now report additional experiments, aiming to gain some empirical understanding of how
the quality of silent interval prediction would affect speech denoising quality.

\begin{table}[ht]
  \caption{Results on how silent interval detection quality affects the speech denoising quality.  
      \label{tab:shift_all}
   }
   \vspace{-2mm}
    \begin{subtable}{.5\linewidth}
      \centering
      \resizebox{0.99\linewidth}{!}{
        \begin{tabular}{cccccc} \toprule
        Shift & No Shift & 1/30 & 1/10 & 1/5 & 1/2 \\ \midrule
        PESQ & \textbf{2.471} & 1.932 & 1.317 & 1.169 & 1.094 \\ \bottomrule
        \end{tabular}
      }
      \caption{Effect of shifting silent intervals.}
      \label{tab:shift}
    \end{subtable}%
    \begin{subtable}{.5\linewidth}
      \centering
        \resizebox{0.99\linewidth}{!}{
        \begin{tabular}{cccccc} \toprule
        Shrink & No Shrink & 20\% & 40\% & 60\% & 80\% \\ \midrule
        PESQ & \textbf{2.471} & 2.361 & 2.333 & 2.283 & 2.249 \\ \bottomrule
        \end{tabular}
        }
        \caption{Effect of shrinking silent intervals.}
        \label{tab:shrink}
    \end{subtable}%
\end{table}

First, starting with ground-truth silent intervals, we shift them on the time axis
by $\nicefrac{1}{30}$, $\nicefrac{1}{10}$, $\nicefrac{1}{5}$, and $\nicefrac{1}{2}$ seconds.
As the shifted time amount increases, more time segments become incorrectly labeled: 
both the numbers of false positive labels (i.e., non-silent time
segments labeled silent) and false negative labels (i.e., silent time
segments are labeled non-silent) increase.
After each shift, we feed the silent interval labels to  
our noise estimation and removal components and measure the denoising quality 
under the PESQ score.

In the second experiment, we again start with ground-truth silent intervals;
but instead of shifting them, we shrink each silent interval toward its center by 
$20\%$, $40\%$, $60\%$, and $80\%$. 
As the silent intervals become more shrunken, fewer time segments are labeled as silent.
In other words, only the number of false negative predictions increases.
Similar to the previous experiment,
after each shrink, we use the silent interval labels in our speech denoising pipeline,
and meausure the PESQ score.



The results of both experiments are reported in \tabref{shift_all}.
As we shrink the silent intervals, the denoising quality drops gently.
In contrast, even a small amount of shift causes a clear drop
of denoising quality.
These results suggest that in comparison to false negative predictions,
false positive predictions affect the denoising quality more negatively.
On the one hand,
reasonably conservative predictions may leave certain silent time segments undetected (i.e., introducing some false negative labels),
but the detected silent intervals indeed reveal the noise profile.
On the other hand,
even a small amount of false positive predictions causes certain non-silent time segments
to be treated as silent segments, and thus the observed noise profile through
the detected silent intervals would be tainted by foreground signals.

\section{Evaluation of Model Performance}



\subsection{Generalization Across Datasets}\label{sec:cross_datasets}
To evaluate the generalization ability of our model, we performed \emph{three}
cross-dataset tests reported in \tabref{cross_dataset}. The experiments are set up in the following way.
\begin{itemize}
    \item ``Test \textbf{i}'': We train our model on our own AVSPEECH+Audioset
        (AA) dataset but evaluate on Valentini's VoiceBank-DEMAND (VD) testset.
        The result is shown in the first row of the ``Test \textbf{i}''
        section in \tabref{cross_dataset}. In comparison, the second row shows the result of training on
        VD and testing on VD.
    \item ``Test \textbf{ii}'': We train our model on our own AA dataset but
        evaluate on our second AVSPEECH+DEMAND (AD) testset. The result is
        shown in the first row of the ``Test \textbf{ii}'' section in \tabref{cross_dataset}. In
        comparison, the second row shows the result of training on AD and
        testing on AD.
    \item ``Test \textbf{iii}'': We train our model on our own AD dataset but evaluate on AA testset. The result is shown in the first row of the ``Test \textbf{iii}'' section of the table. In comparison, the second row shows the result of training on AA and testing on AA.
\end{itemize}
The small degradation in each cross-dataset test demonstrates the great generalization ability of our method.
We could not directly compare the generalization ability of our model with
existing methods, as no previous work reported cross-dataset evaluation
results.

\begin{table}[ht]
\begin{tabular}{lccccccc} \toprule
Test & Trainset & Testset & PESQ  & CSIG  & CBAK  & COVL  & STOI\\ \midrule
\multirow{2}{*}{Test \textbf{i}} & AA & VD & 3.00 & 3.78 & 3.08 & 3.34 & 0.98 \\
 & VD & VD & 3.16 & 3.96 & 3.54 & 3.53 & 0.98 \\ \midrule
\multirow{2}{*}{Test \textbf{ii}} & AA & AD & 2.65 & 3.48 & 3.21 & 3.01 & 0.90 \\
 & AD & AD & 2.80 & 3.66 & 3.36 & 3.17 & 0.91 \\ \midrule
\multirow{2}{*}{Test \textbf{iii}} & AD & AA & 2.12 & 2.71 & 2.65 & 2.34 & 0.79 \\
 & AA & AA & 2.30 & 2.91 & 2.81 & 2.54 & 0.82 \\ \bottomrule
\end{tabular}
\caption{Generalization across datasets.}
\label{tab:cross_dataset}
\end{table}

\subsection{Generalization on Real-world Data}\label{sec:rw}

We conduct experiments to understand the extent to which the model trained with 
different datasets can generalize to real-world data.
We train two versions of our model using
our AVSPEECH+Audioset dataset and Valentini's VoiceBank-DEMAND, respectively
(denoted as ``Model AA'' and ``Model VD'', respectively, in \tabref{rw_data}), and use them to
denoise our collected real-world recordings. For the denoised real-world
audios, we measure the noise level reduction in detected silent intervals. This
measurement is doable, since it requires no knowledge of noise-free
ground-truth audios. As shown in~\tabref{rw_data}, in terms of noise reduction,
the model trained with our own AA dataset outperforms the one trained with the
public VoiceBank-DEMAND dataset by a significant amount for \textit{all} tested
real-world recordings.  On average, it produces \textbf{22.3 dB} noise
reduction in comparison to \textbf{12.6 dB}, suggesting that our dataset allows
the denoising model to better generalize to many real-world scenarios.

\begin{table}[ht]
\begin{tabular}{lcc} \toprule
Real-world Recording & Model AA noise reduction (dB) & Model VD noise reduction (dB) \\ \midrule
Song Excerpt 1 & \textbf{22.38} & 9.22 \\
Song Excerpt 2 & \textbf{21.16} & 17.65 \\
Chinese & \textbf{18.99} & 15.65 \\
Japanese & \textbf{16.95} & 9.39 \\
Korean & \textbf{25.90} & 9.76 \\
German & \textbf{14.54} & 7.29 \\
French & \textbf{21.95} & 17.16 \\
Spanish 1 & \textbf{33.34} & 11.00 \\
Spanish 2 & \textbf{31.66} & 14.09 \\
Female 1 & \textbf{17.64} & 7.79 \\
Female 2 & \textbf{30.21} & 8.64 \\
Female 3 & \textbf{19.15} & 6.70 \\
Male 1 & \textbf{24.81} & 14.44 \\
Male 2 & \textbf{24.92} & 13.35 \\
Male 3 & \textbf{13.10} & 10.99 \\
Multi-person 1 & \textbf{32.00} & 21.60 \\
Multi-person 2 & \textbf{15.10} & 15.07 \\
Multi-person 3 & \textbf{18.71} & 10.68 \\
Street Interview & \textbf{20.54} & 18.94 \\ \midrule
AVERAGE & \textbf{22.27} & 12.60 \\ \bottomrule
\end{tabular}
\caption{Real-world recording noise reduction comparison.}
\label{tab:rw_data}
\end{table}
